\documentclass[preprint]{aastex}

\usepackage{natbib}
\shorttitle{XX Cep} 
\shortauthors{J.-R. KOO ET AL.} 

\begin{document}
\title{Time-series spectroscopy of the pulsating eclipsing binary XX Cephei}
\author{Jae-Rim Koo$^{1}$, Jae Woo Lee$^{1,2}$, Kyeongsoo Hong$^{1}$, Seung-Lee Kim$^{1,2}$, and Chung-Uk Lee$^{1,2}$}%, \\
\affil{$^{1}$Korea Astronomy and Space Science Institute, Daejeon (34055), Republic of Korea;\email{koojr@kasi.re.kr}}
\affil{$^{2}$Astronomy and Space Science Major, University of Science and Technology, Daejeon (34113), Republic of Korea}
%\affil{$^{2}$University of Science and Technology, Daejeon (34113), Republic of Korea}

%\maketitle

\begin{abstract}
Oscillating Algol-type eclipsing binaries (oEA) are very interesting objects that have three observational features of eclipse, pulsation, and mass transfer. 
Direct measurement of their masses and radii from the double-lined radial velocity data and photometric light curves would be the most essential 
for understanding their evolutionary process and for performing the asteroseismological study.
We present the physical properties of the oEA star XX Cep from high-resolution time-series spectroscopic data.
The effective temperature of the primary star was determined to be 7,946 $\pm$ 240 K by comparing the observed spectra and the Kurucz models. 
We detected the absorption lines of the secondary star, which had never been detected in previous studies, and obtained the radial velocities for both components.
With the published $BVRI$ light curves, we determined the absolute parameters for the binary via Wilson-Devinney modeling.
The masses and radii are $M_{1} = 2.49 \pm 0.06$ $M_\odot$, $M_{2} = 0.38 \pm 0.01$ $M_\odot$, $R_{1} = 2.27 \pm 0.02$ $R_\odot$, and $R_{2} = 2.43 \pm 0.02$ $R_\odot$, respectively. 
The primary star is about $45 \%$ more massive and $60 \%$ larger than the zero-age main sequence (ZAMS) stars with the same effective temperature.
It is probably because XX Cep has experienced a very different evolutionary process due to mass transfer, contrasting with the normal main sequence stars.
The primary star is located inside the theoretical instability strip of $\delta$ Sct-type stars on HR diagram. 
We demonstrated that XX Cep is an oEA star, consisting of a $\delta$ Sct-type pulsating primary component and an evolved secondary companion. 

\end{abstract}

\section{Introduction}
Eclipsing binaries offer the chance for direct measurement of physical parameters such as mass, size, and temperature.
The accurate determination of these quantities is possible for double-lined eclipsing binaries with both light and radial velocity (RV) curves.
We could understand the structure and evolution of the eclipsing binaries from these physical properties.
For close binaries, it is not easy to separate the binary components in the observed spectra because they are blended. 
Moreover, it is so hard to detect the features of a secondary component with a very small light contribution to the binary systems, as in the case of classical Algol-type eclipsing binaries.

Meanwhile, the oscillating Algol-type eclipsing binaries \citep[oEA stars;][]{mkrtichian2004} have a mass-accreting pulsating star as a primary component.
Pulsating stars in these systems are interesting objects for asteroseismology and for the study of stellar interiors and evolution because the binary provides useful information about the components.
The mass-accretion on the surface of pulsating gainers is an important feature of oEA stars, compared to classical $\delta$ Sct-type stars in the detached systems. 
We could study the short-term dynamical processes such as the acceleration and braking of stellar rotation and the evolution of systems through this feature.
However, only eleven oEA stars were studied using double-lined RV curves with/without light curves 
(\citealp{hoffman2009,tkachenko2009,tkachenko2010,lampens2011,soydugan2013,hong2015}).
We studied the oEA star Y Cam as the first target \citep{hong2015}. They obtained spectroscopic signals from each component star and presented the absolute dimensions and evolutionary status of the system, using previously published light curves and their own RV curves.

We performed a time-series spectroscopic study for the oEA star candidate XX Cep (RA$_{\rm{J}2000.0}$ $= 23^h 38^m 20^s.29$, DEC$_{\rm{J}2000.0}$ $= +64\arcdeg20\arcmin02\arcsec.7$; $V = 9.20$, $B-V = 0.236$). 
Since the discovery of XX Cep by \citet{gengler1928}, many studies of photometric and spectroscopic observations have been performed. 
\citet{struve1946} measured the RVs for the primary star and determined its spectral type to be about A8. 
From the period analyses, mass transfer between the components and/or circumbinary companions around the eclipsing pair were expected.
\citet{angione2006} determined a spectral type of the primary star to be A4 from their spectroscopic observations, 
but features were not detected for the secondary star or mass transfer. 
Recently, the primary component was proposed to be a $\delta$ Sct-type pulsating star with a frequency of 31.769 cd$^{-1}$ by \citet{lee2007}, and \citet{hosseinzadeh2014} confirmed the pulsating frequency. 
In this paper, we present the precise physical properties and evolutionary state of XX Cep from detailed analyses of our high-resolution time-series spectroscopic data, along with previously published $BVRI$ light curves.

\section{Observations and Data Analyses}
We performed time-series spectroscopic observations for XX Cep to find signatures of the secondary star and to determine the absolute parameters of each component. 
High-resolution spectra for XX Cep were obtained for six nights between October 2007 and November 2008
using a fiber-fed spectrograph BOES \citep[Bohyunsan Optical Echelle Spectrograph,][]{kim2007}
attached to the 1.8 m reflector at BOAO (Bohyunsan Optical Astronomy Observatory), Korea. 
We used the largest fiber with a diameter of 300 $\mu$m, corresponding to a field of view of 4.3 arcsec and a resolving power of 30,000.
Twenty-six spectra were obtained, with exposure times of 3,600 seconds. Th-Ar Arc images for wavelength calibration and Tungsten-Halogen lamp images for flat correction were also taken before and after the observations.
The signal-to-noise ratio (SNR) at around $5550~{\rm \AA}$ was about 80.

\begin{figure}
\includegraphics[]{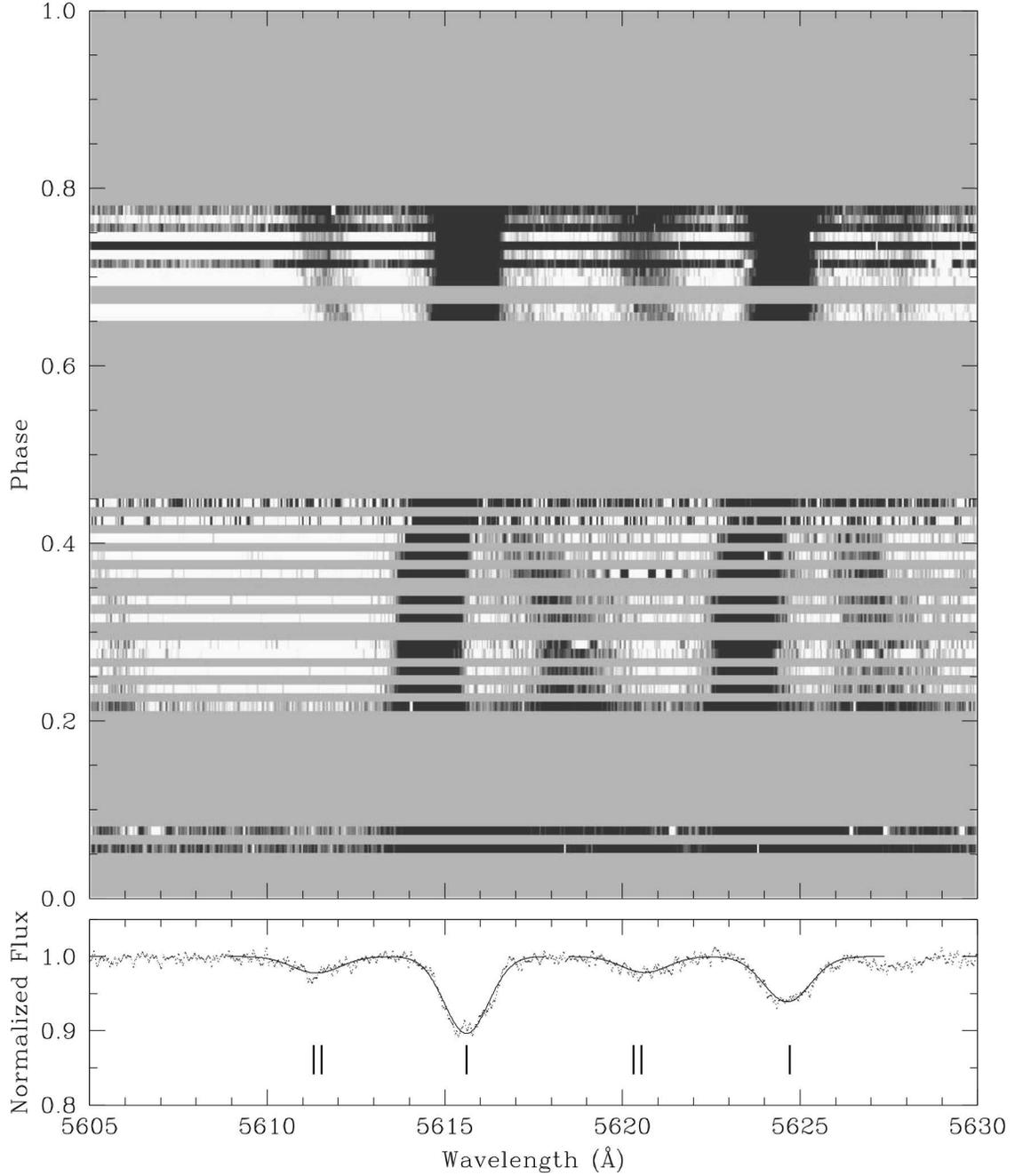}
\caption{Trailed spectra and Gaussian fitting sample for RVs in the regions of Fe I $\lambda$5615.64 and $\lambda$5624.54. In upper panel, 
two components can be identified easily, shifted through orbital phases. In lower panel, dots and lines represent the observations and Gaussian fittings at phase of $\phi = 0.75$ , respectively. `$|$' for primary and `$||$' for secondary components are marked.
\label{Fig1}}
\end{figure}

To obtain accurate physical properties for an eclipsing binary, RV curves for both components and light curves are required. 
\citet{struve1946} and \citet{angione2006} obtained the spectroscopic data but they could not find any evidence for a secondary component due to its low light contribution and poor spectral resolution. 
However, we were able to find three isolated double lines (Fe I $\lambda5615.64$, Fe I $\lambda5624.54$, and Ca I $\lambda6102.72$) in our high-resolution spectra. 
The RVs of each component were measured using the line-profile fitting with two Gaussian functions, in a way similar to that of \citet{koo2014}. Figure \ref{Fig1} displays the trailed spectra of Fe I $\lambda5615.64$ and $\lambda5624.54$ regions, and a sample of RV fittings for the phase of $\phi = 0.75$. 
The averaged RVs and the standard deviations of each component are listed in Table \ref{tab1}.

\begin{figure}
\includegraphics[]{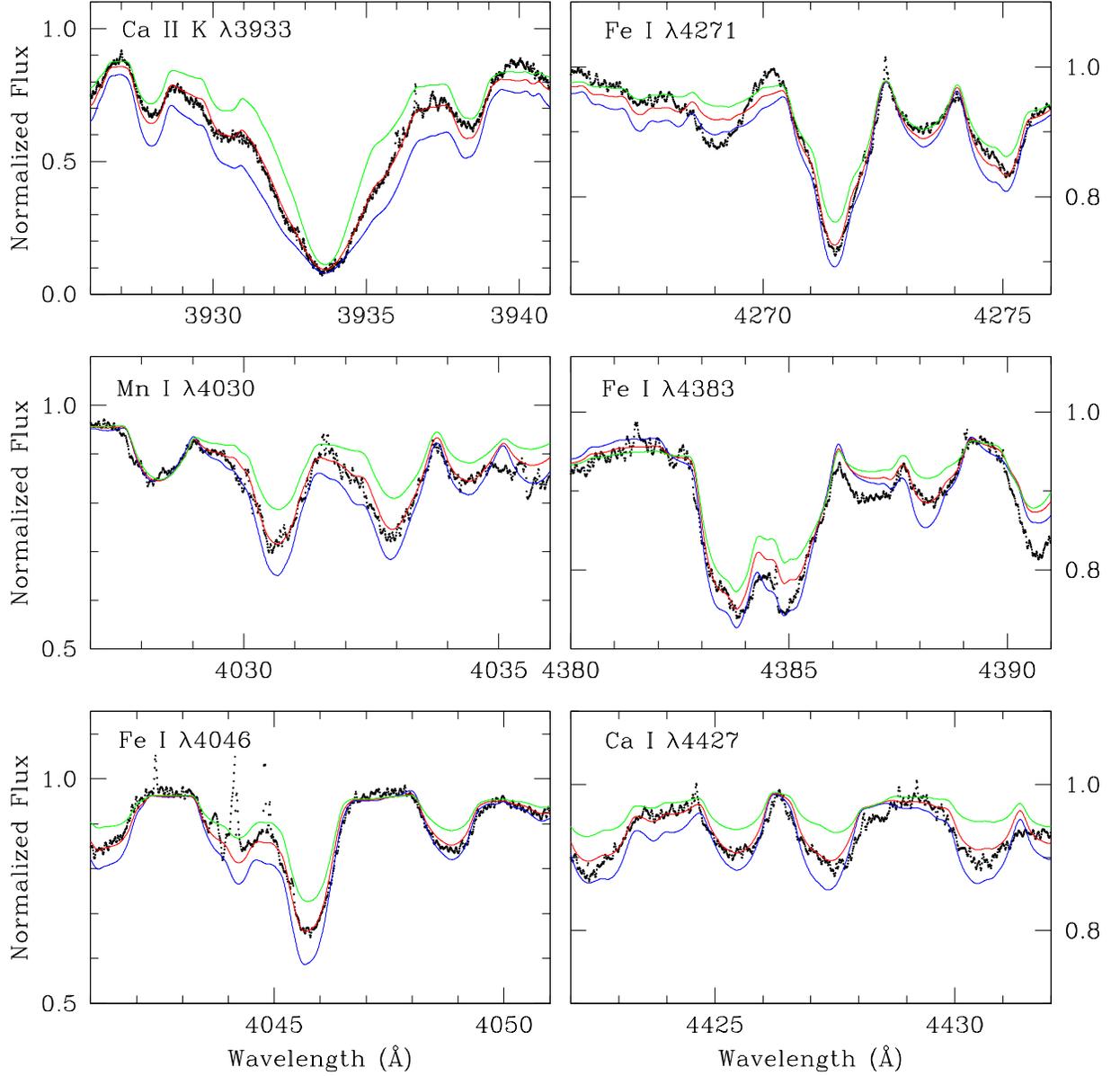}
\caption{Six spectral regions of XX Cep. Dots represent the combined spectrum and blue, red, and green lines represent the synthetic spectra of $T_{\rm eff} = 7,500$ K, 8,000 K, and 8,500 K with solar abundance, respectively.
\label{Fig2}}
\end{figure}

\begin{figure}
\includegraphics[]{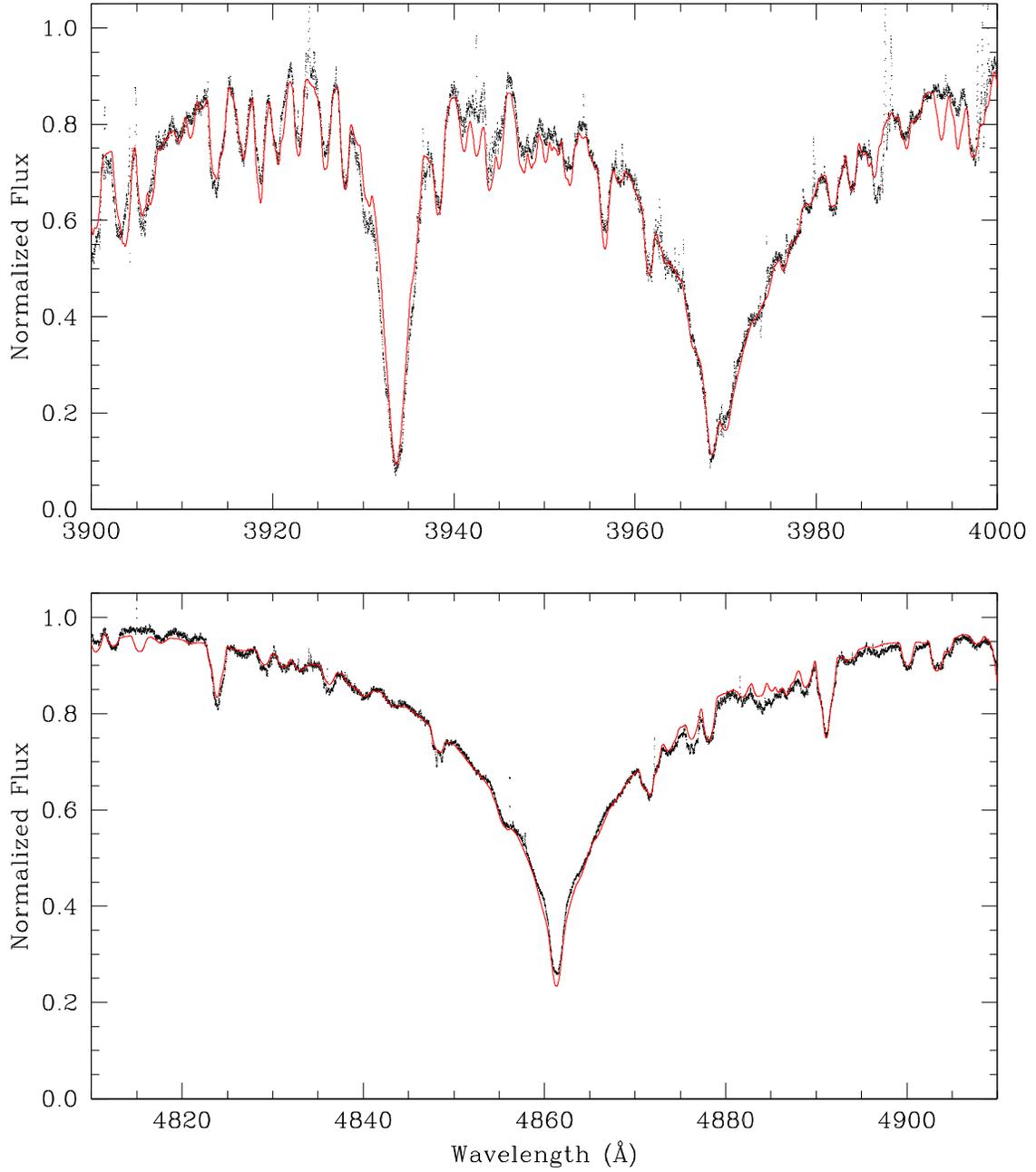}
\caption{Combined spectrum of XX Cep (dots) and the synthetic model of $T_{\rm eff} = 7,946$ K (lines). Upper and lower panels show the Ca II H \& K and H$\beta$ regions, respectively. 
\label{Fig3}}
\end{figure}

\citet{angione2006} classified the spectral type of the primary as A4V by comparing the width of the H$\beta$ line with those of other standard star's.
However, it was not consistent with the earlier result (A8) by \citet{struve1946}. Therefore, we tried to determine the spectral type of the hotter star.
To obtain a high SNR spectrum of the primary star, we selected spectra with different phases and then shifted them to zero velocities 
considering the RVs of the primary component. A final spectrum was obtained by applying a median combine. 
The effective temperature of the primary star was derived by comparing the combined spectrum with the synthetic model grids by \citet{castelli2004}. 
We chose several useful lines in the temperature classification of A-type stars as noted in the $A~Digital~Spectral~Classification~Atlas$ by R. O. Gray. 
Figure \ref{Fig2} shows the six regions of observed and synthetic spectra of $T_{\rm eff} = 7,500$ K, 8,000 K, and 8,500 K with solar abundance. 
The synthetic spectrum of $T_{\rm{eff}}$ $\sim$ 8,000 K was best fitted to our data. 
To check and improve the effective temperature, iSpec (the integrated spectroscopic framework) by \citet{blanco2014} was examined.
We adopted the ATLAS9 model atmosphere \citep{castelli2004} and assumed solar abundance.
Because strong hydrogen and Ca II K lines can be used to determine the effective temperature for the A-type stars, we chose the regions of H$\beta$ and Ca II H \& K. The effective temperature was determined to be $7,946 \pm 240$ K, which is in very good agreement with the above value of $\sim$ 8,000 K.
Figure \ref{Fig3} displays the two spectral regions of our spectrum and the synthetic one of $T_{\rm eff} = 7,946$ K. 
Thus, we classified the spectral type of the primary star to be A6V.
Our temperature of the primary star is cooler than $T_{\rm eff} = 8,500$ K given by \citet{angione2006}, but is in good agreement with empirical relations between temperature and color index, considering the de-reddened color index of $(B-V)_{0} = 0.192$ derived by applying 3D modeling of the Galaxy by \citet{drimmel2003}.

We applied iSpec to determine the rotational velocity of the primary star. Effective temperature and surface gravity were fixed to be $T_{\rm eff} = 7,946$ K 
and log $g = 4.12$ from binary solutions obtained in the following section. We examined the metallic line region between $5000~\rm\AA$ and $5800~\rm\AA$, 
and determined a rotational velocity of $v_{\rm{1}}$sin$i = 48.6 \pm 6.8$ km s$^{-1}$. 
Within errors, this value is consistent with $47 \pm 2$ km s$^{-1}$ from \citet{etzel1993} and corresponds to calculated synchronous rotation velocity of $v_{\rm 1,sync} = 49.2 \pm 0.4$ km s$^{-1}$.

\section{Binary Modeling and Absolute Dimensions}

To obtain the binary solution of XX Cep, our RV curves were analyzed with the $BVRI$ light curves obtained in 2002 and 2003 by \citet{lee2007} 
using the 2013 version of the Wilson-Devinney synthesis code \citep[][hereafter WD]{wilson1971,wilson2014}. 
Because XX Cep had already been analyzed in several photometric studies, and is known to be a semi-detached eclipsing binary where the secondary star fills its inner Roche lobe, we adopted mode 5 in the WD synthesis. 
Our syntheses were performed in two stages, following the procedure described by \citet{hong2015}. 
First, we analyzed the RV curve alone with the photometric solutions of \citet{lee2007}. 
Then the light curves were solved using the spectroscopic parameters obtained in the first stage. 
These steps were repeated until the results were consistent with each other.
The values with parenthesized errors in Tables \ref{tab2} and \ref{tab3} signify adjusted parameters.
In this paper, the subscripts 1 and 2 refer to the primary and secondary stars, respectively.

The effective temperature of the primary star was fixed to be $T_{1} = 7,946$ K from our spectral analyses. 
We assumed the gravity-darkening exponents and bolometric albedoes for each component to be $g_{1} = 1.0$, $A_{1} = 1.0$ and $g_{2} = 0.32$, $A_{2} = 0.5$ 
which are appropriate for stars with radiative and convective envelopes, respectively. 
The square root bolometric ($X$, $Y$) and 
monochromatic ($x$, $y$) limb-darkening coefficients were interpolated from the values of \citet{van1993} in concert with 
the model atmosphere option. 
The projected rotational velocity of $v_1$sin$i = 48.6 \pm 6.8$ km s$^{-1}$ for the primary star is in good agreement with a synchronized rotation velocity of $v_{1,\rm sync} = 49.2 \pm 0.4$ km s$^{-1}$ within their errors. We can assume that the system is tidally locked in synchronous rotation. 
Therefore, we set the rotation parameters $F_1 = F_2 = 1.0$ in the WD modelings.

\begin{figure}
\includegraphics[]{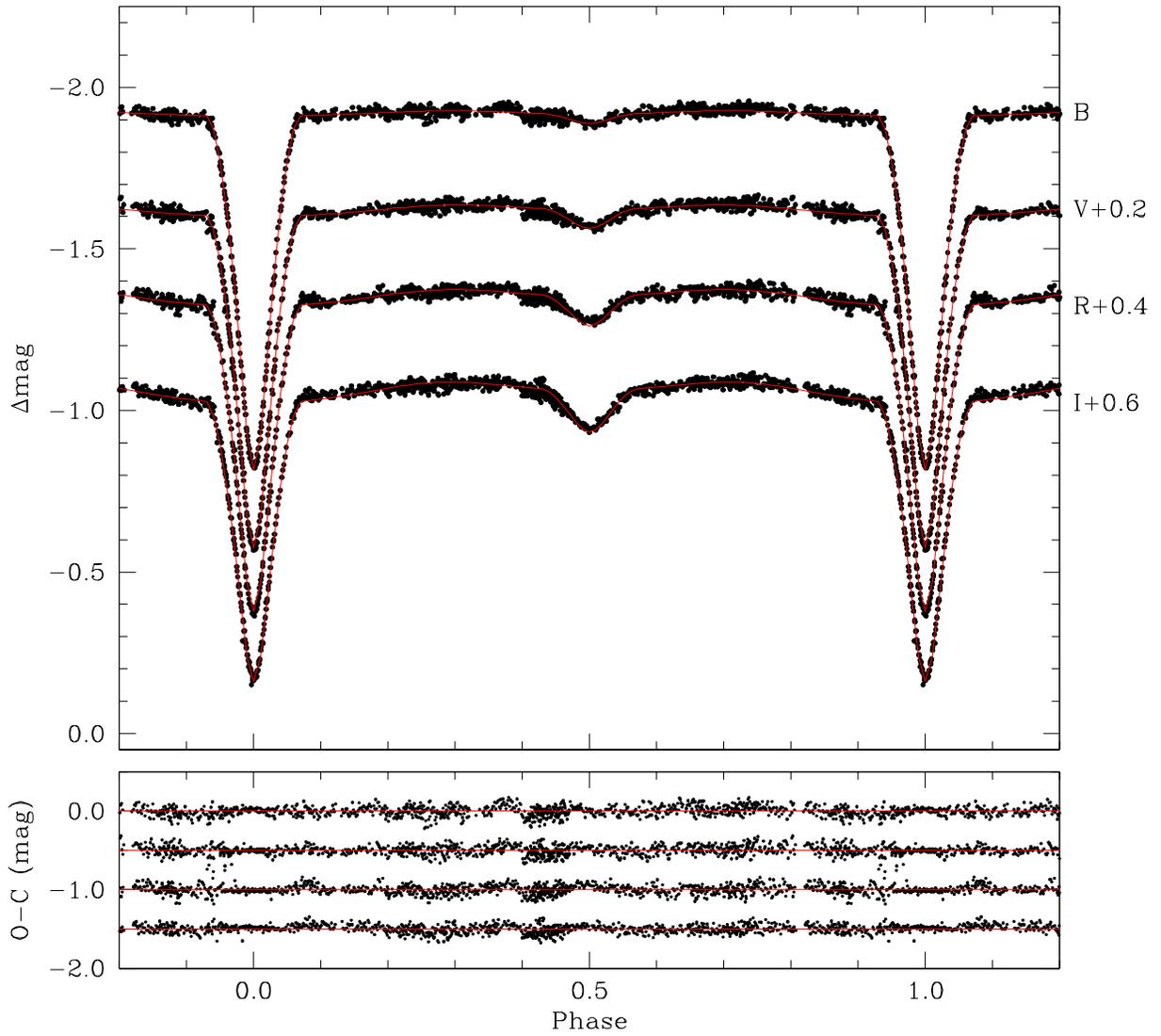}
\caption{$BVRI$ light curves of XX Cep with fitted models. Measurements (black dots) were taken from \citet{lee2007}. Lower panel displays the differences between measurements and models.
\label{Fig4}}
\end{figure}

\begin{figure}
\includegraphics[]{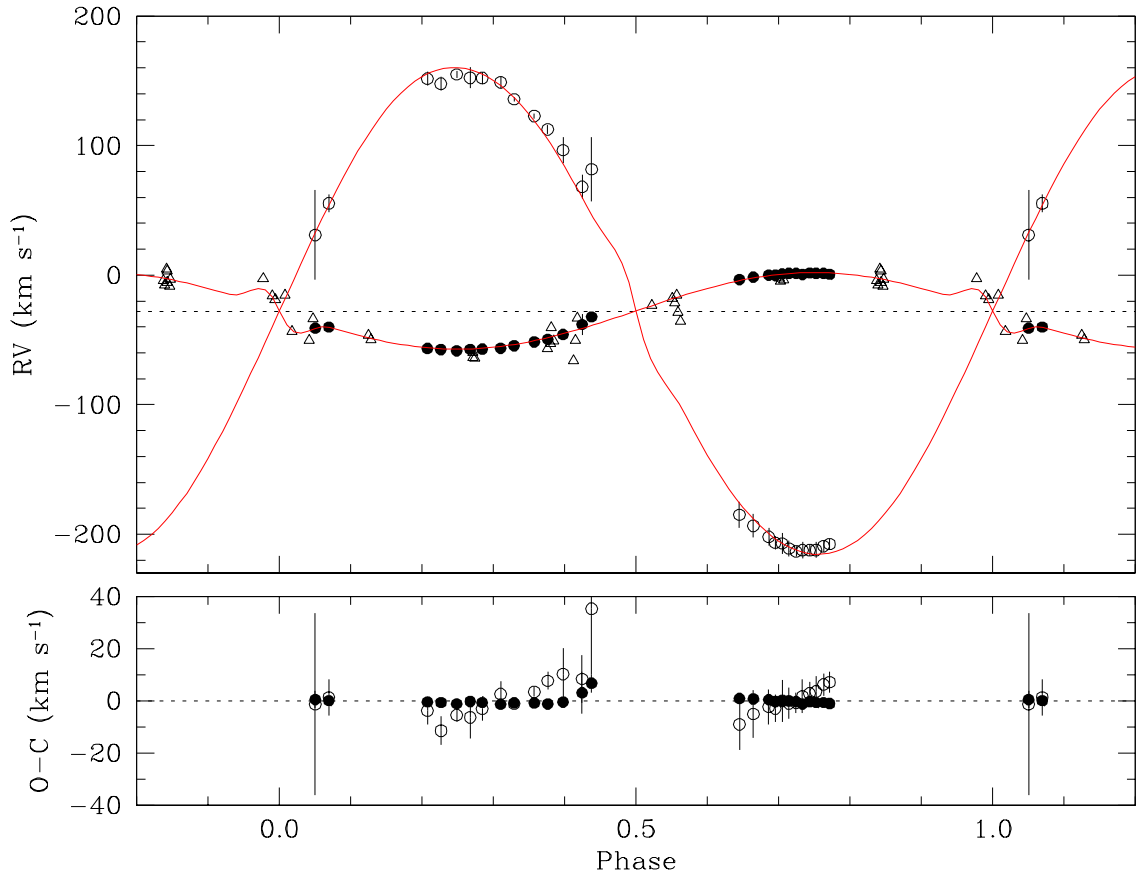}
\caption{RV curves of XX Cep with fitted models. Filled and open circles represent the RV measurements for the primary and secondary components, respectively. Earlier observations by \citet{struve1946} were plotted with open triangles. The dotted line refers to the system velocity of $-27.57$ km s$^{-1}$ in the upper panel. Residuals between observations and models are displayed in the lower panel.
\label{Fig5}}
\end{figure}

The orbital period of XX Cep has varied in complex ways due to a combination of causes rather than in a monotonic fashion, such as a parabola or a sinusoid \citep{lee2007, hosseinzadeh2014}.
Because the time gap between photometric and 
spectroscopic observations is about 5 years, we calculated the epoch ($T_0$) and period ($P$) for the spectroscopic data with WD code to be 2,454,397.1139 (87) and 
2.337357 (71) days, respectively. Then spectroscopic parameters such as system velocity ($\gamma$), semi-major axis ($a$), and mass ratio ($q$) were adjusted in the WD analysis 
and listed in Table \ref{tab2}. The spectroscopic mass ratio $q$ is smaller than previous results from analysis of only the photometric data.
In the light-curve analysis, the ephemeris was fixed to be values of \citet{lee2007}.
The final solutions were listed in Table \ref{tab3}, where $r$(volume) is the mean volume radius calculated from the tables of \citet{mochnacki1984}.
Figures \ref{Fig4} and \ref{Fig5} display the light and RV curves with fitted models, respectively.

From our light and RV solutions, the absolute dimensions for each component were calculated using the JKTABSDIM code \citep{southworth2005} and listed in Table \ref{tab4}.
The luminosity ($L$) and bolometric magnitude were computed by adopting $T_{\rm{eff}\odot} = 5,780$ K and $M_{\rm{bol}\odot} = +4.73$ for solar values. 
The bolometric corrections (BCs) were obtained from the relation between $\log T_{\rm eff}$ and BC given by \citet{torres2010}. 
The mass of the primary star is much heavier than normal main-sequence stars with the same effective temperature but is consistent with the mass-luminosity relation of $L_1 \propto M^{3.20} $ for the semi-detached binaries given by \citet{ibanoglu2006}. 
This may be because XX Cep has experienced evolutionary processes different from those of single stars through mass transfer from the secondary to the primary star.

Since XX Cep is located near the Galactic plane (Galactic latitude of $b = 2.58$ deg), we adopted the Galactic 3D model \citep{drimmel2003} for interstellar reddening. 
With the apparent visual magnitude ($V = 9.20$) at quadrature from \citet{bonifazi1986}, and the interstellar reddening of $A_{\rm V} = 0.136$ adopting the trigonometric parallaxes of 4.08 mas \citep[][HIP116648]{leeuwen2007}, the distance to the system was calculated to be $329 \pm 23$ pc. 
This is marginally consistent with the distance of $245 \pm 61$ pc taken by the parallax, considering their errors.

\section{Summary and Conclusions}
In this paper, we presented the precise physical properties of the oEA star candidate XX Cep from new high-resolution spectroscopic data and previously published light curves. 
The absorption lines from the primary and secondary components were detected, which had not previously been done for the secondary in previous studies, 
and the RV curves of each component were obtained. The effective temperatures of both stars were determined to be $T_{1} = 7,946$ K and $T_{2} = 4,483$ K 
from our spectral analysis and WD modeling, respectively. From the simultaneous analyses of the RV and light curves, accurate absolute dimensions of the binary system were 
calculated as follows: $M_{1} = 2.49 \pm 0.06$ $M_\odot$, $M_{2} = 0.38 \pm 0.01$ $M_\odot$, $R_{1} = 2.27 \pm 0.02$ $R_\odot$, $R_{2} = 2.43 \pm 0.02$ $R_\odot$, 
$L_1 = 18.4 \pm 2.3$ $L_\odot$, and $L_2 = 2.1 \pm 0.5$ $L_\odot$. 
The effective temperature and mass for the primary star are very different compared to previous results ($T_{1} = 8,500$ K and $M_{1} = 1.92$ $M_\odot$) from \citet{angione2006}. 
However, we believe that our results are more reliable, because the temperature presented in this paper were measured from the higher resolution spectra and the other parameters such as mass and radius were directly determined from both double-lined RVs and multiband light curves.
Thus, we conclude that XX Cep consists of a massive primary star of spectral type A6V and a secondary component of  spectral type K3IV.

\begin{figure}
\includegraphics[height=0.9\textheight]{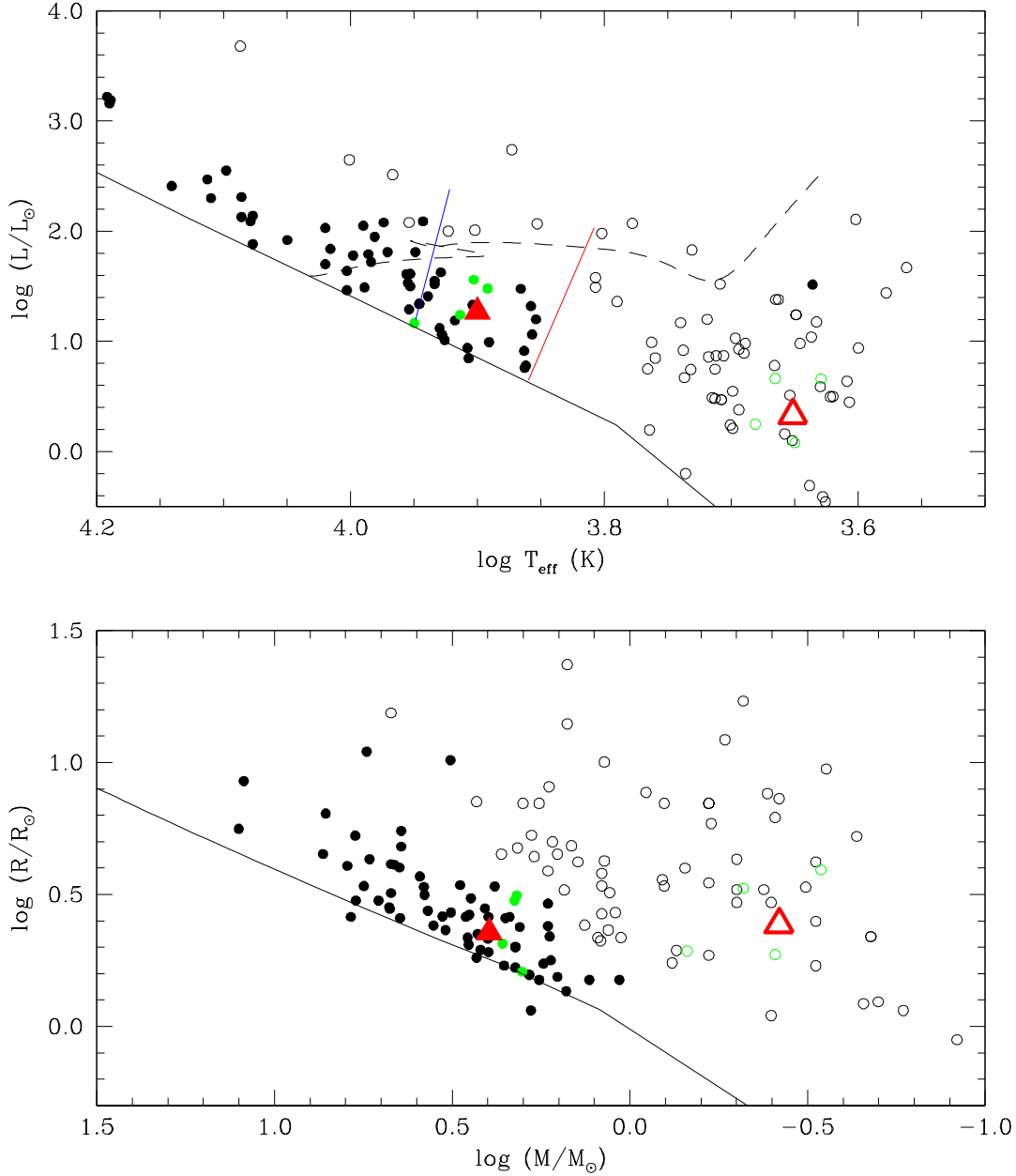}
\caption{The position on the HR and mass-radius diagrams for XX Cep (triangles) and other semi-detached Algol-type eclipsing binaries (circles) by \citet{ibanoglu2006}. Filled and open symbols refer to the primary and secondary stars, respectively. Green circles indicate well-studied oEA stars with double-lined RV curves. The black solid line displays the ZAMS calculated using equations adopted from \cite{tout1996} of Z $= 0.02$. In the upper panel, blue and red lines represent the $\delta$ Sct instability strips, and
the dashed-line denotes the evolutionary track of a normal main-sequence star having the mass of $2.5$ $M_\odot$ and solar abundance.
\label{Fig6}}
\end{figure}

Figure \ref{Fig6} displays HR and mass-radius diagrams for XX Cep, and for other semi-detached Algol-type eclipsing binaries taken from the compilations of \citet{ibanoglu2006}.
Green circles represent the well-studied oEA stars with RV curves of both components.
The primary component of XX Cep is located near the main-sequence, while the secondary one has larger size considering its mass. 
These are normal characteristics of an Algol-type eclipsing binary as shown in Figure \ref{Fig6}, and can be explained from the evolutionary stage of binary systems.
Initial more massive star fills its Roche-lobe and a mass transfer may be occurred to the less massive star via Lagrange L$_{1}$ point. 
As a result, the more massive star loses most of its own mass and becomes the present secondary star, but remains in the Roche-lobe filled stage. 
No emission feature of the mass transfer was detected in our data, as in the previous study by \citet{angione2006}. 

In previous photometric studies, the light curves of XX Cep showed a pulsating feature with a frequency of about 32 cd$^{-1}$ \citep{lee2007,liakos2012,hosseinzadeh2014} and the position of the primary star of XX Cep on the HR diagram is also inside the $\delta$ Sct instability strip.
 We calculated the pulsation constant $Q$ using the formula of \citet{petersen1972}:
\begin{displaymath}
{\rm log}~Q = -6.454+0.5~{\rm log}~g+0.1~M_{\rm bol}+{\rm log}~\it{T}_{\rm eff}+{\rm log}~P_{\rm pul},
\end{displaymath}
where $P_{\rm{pul}}$ is the pulsating period.
The calculated value of $Q = 0.0144$ could be identified as fourth overtone of p-mode pulsations \citep{fitch1981}.
Further, the position of the primary star of XX Cep on the HR diagram is also inside the $\delta$ Sct instability strip. 
Therefore, we can demonstrate that the primary star of the semi-detached binary XX Cep is a $\delta$ Sct-type pulsating star, and thus the binary is an oEA star. 
The oEA stars are very important systems for understanding the stellar interiors and for asteroseismology, because we can determine the precise physical parameters of double-lined eclipsing binaries. 
However, only handful systems have been well studied because it is difficult to detect the faint secondary components in the observed spectra. 
We will continue the studies of these objects to determine accurately their physical properties from high resolution time-series spectroscopic observations.

\acknowledgments{
This work was supported by KASI (Korea Astronomy and Space Science Institute) grant 2016-1-832-01. 
We have used the Simbad database maintained at CDS, Strasbourg, France.

}

\clearpage

\begin{deluxetable}{lrr}
\tabletypesize{\scriptsize}
\tablewidth{0pt}
\tablecaption{Radial velocities for XX Cep
\label{tab1}}
\tablehead{
\colhead{HJD}   & \colhead{$V_1$ }            &   \colhead{$V_2$}    \\
\colhead{}      & \colhead{(km s$^{-1}$)}     & \colhead{(km s$^{-1}$)}
}
\startdata
2454397.2311 &  $-$40.9 $\pm$    3.3 &   31.1   $\pm$   34.8   \\
2454397.2759 &  $-$40.1 $\pm$    0.8 &   55.4   $\pm$    7.0   \\
2454398.1364 &  $-$32.2 $\pm$    3.5 &   81.7   $\pm$   25.0   \\
2454401.0759 &   $-$0.1 $\pm$    1.4 & $-$206.6 $\pm$    5.2   \\
2454401.1207 &    1.3   $\pm$    2.2 & $-$211.1 $\pm$    6.0   \\
2454401.1650 &    0.6   $\pm$    0.7 & $-$212.3 $\pm$    6.5   \\
2454401.2094 &    1.3   $\pm$    1.9 & $-$212.2 $\pm$    6.0   \\
2454401.2537 &    0.7   $\pm$    1.1 & $-$207.6 $\pm$    4.1   \\
2454771.9257 &  $-$51.6 $\pm$    1.2 &  122.8   $\pm$    1.8   \\
2454771.9702 &  $-$49.6 $\pm$    0.8 &  112.6   $\pm$    3.4   \\
2454772.0202 &  $-$45.9 $\pm$    1.8 &   96.5   $\pm$   10.0   \\
2454772.0826 &  $-$38.1 $\pm$    8.1 &   68.2   $\pm$    9.1   \\
2454773.9133 &  $-$56.5 $\pm$    0.6 &  151.7   $\pm$    5.3   \\
2454773.9578 &  $-$57.5 $\pm$    1.8 &  147.8   $\pm$    5.5   \\
2454774.0091 &  $-$58.3 $\pm$    2.1 &  155.0   $\pm$    2.7   \\
2454774.0535 &  $-$57.3 $\pm$    1.7 &  152.4   $\pm$    8.0   \\
2454774.0930 &  $-$57.2 $\pm$    2.0 &  152.2   $\pm$    4.7   \\
2454774.1531 &  $-$56.4 $\pm$    1.3 &  148.8   $\pm$    5.0   \\
2454774.1976 &  $-$54.4 $\pm$    1.9 &  135.8   $\pm$    1.6   \\
2454774.9356 &   $-$3.4 $\pm$    1.2 & $-$185.0 $\pm$    9.9   \\
2454774.9810 &   $-$1.5 $\pm$    1.7 & $-$193.5 $\pm$    9.0   \\
2454775.0313 &    0.0   $\pm$    1.7 & $-$202.0 $\pm$    6.7   \\
2454775.0756 &    0.8   $\pm$    2.3 & $-$207.2 $\pm$    8.0   \\
2454775.1212 &    1.2   $\pm$    1.7 & $-$213.3 $\pm$    4.0   \\
2454775.1654 &    1.6   $\pm$    1.7 & $-$212.4 $\pm$    4.4   \\
2454775.2110 &    1.2   $\pm$    1.1 & $-$209.3 $\pm$    4.3   \\
              
\enddata                                      
\end{deluxetable}

\begin{deluxetable}{lc}
\tabletypesize{\scriptsize}
\tablewidth{0pt}
\tablecaption{Spectroscopic elements of XX Cep
\label{tab2}}
\tablehead{
\colhead{Parameter}   & \colhead{Value}
}
\startdata
$T_0$ ($\rm HJD$)	 & 2,454,397.1139(87) \\
$P$					 & 2.337357(71) \\
$K_1$ (km s$^{-1}$)	 & 29.6(0.3) \\
$K_2$ (km s$^{-1}$)	 & 196.2(1.8) \\
$\gamma$ (km s$^{-1}$) &  $-$27.57(34)    \\
\enddata                                      
\end{deluxetable}

\begin{deluxetable}{lcc}
\tabletypesize{\scriptsize}
\tablewidth{0pt}
\tablecaption{Binary parameters of XX Cep
\label{tab3}}
\tablehead{
\colhead{Parameter}   & \colhead{Primary} & \colhead{Secondary}
}
\startdata
$a$ (R$_\odot$)      &    \multicolumn{2}{c}{10.53(10)}        \\
$q$                  &    \multicolumn{2}{c}{0.151(2)}         \\
$i$ (deg)            &    \multicolumn{2}{c}{82.37(2)}         \\
$T$ (K)              &      7,946         &    4,483(9)        \\
$\Omega$             &      4.805(13)    &    2.106            \\
$x_{B}$, $y_{B}$     &      0.790, 0.329  &   0.851, $-$0.154   \\
$x_{V}$, $y_{V}$     &      0.685, 0.299  &   0.813,  0.014     \\
$x_{R}$, $y_{R}$     &      0.569, 0.278  &   0.737,  0.116     \\
$x_{I}$, $y_{I}$     &      0.464, 0.255  &   0.641,  0.162     \\
$l$/($l_1+l_2$)$_B$  &      0.963(2)     &    0.037            \\
$l$/($l_1+l_2$)$_V$  &      0.919(4)     &    0.081            \\
$l$/($l_1+l_2$)$_R$  &      0.868(5)     &    0.132            \\
$l$/($l_1+l_2$)$_I$  &      0.815(5)     &    0.185            \\
$r$ (pole)           &      0.2148(6)    &   0.2146            \\
$r$ (point)          &      0.2167(6)    &   0.3167            \\
$r$ (side)           &      0.2161(6)    &   0.2232            \\
$r$ (back)           &      0.2165(6)    &   0.2551            \\
$r$ (volume)$\rm ^a$ &    0.2159          &  0.2314             \\

\enddata                                      
\tablenotetext{a}{Mean volume radius.}
\end{deluxetable}

\begin{deluxetable}{lcc}
\tabletypesize{\scriptsize}
\tablewidth{0pt}
\tablecaption{Physical properties for XX Cep
\label{tab4}}
\tablehead{
\colhead{Parameter}   & \colhead{Primary} & \colhead{Secondary}
}
\startdata
Mass (M$_\odot$)   &       2.49(6)     &    0.38(1)            \\
Radius (R$_\odot$) &       2.27(2)     &    2.43(2)           \\
log $g$ (cgs)      &      4.121(5)     &   3.240(5)         \\
log $L$ (L$_\odot$) &      1.27(5)     &    0.33(9)           \\
$M_{\rm bol}$ (mag) &      1.57(13)     &   3.91(24)            \\
BC (mag)           &      0.02          &   $-$0.61          \\
$M_V$ (mag)        &      1.55(13)     &   4.52(24)        \\
Distance (pc)      &        \multicolumn{2}{c}{329(23)}     \\

\enddata                                      
\end{deluxetable}

\clearpage
  

\end{document}